\documentclass[lettersize, journal]{IEEEtran}
\usepackage{amsmath,amsfonts}
\usepackage{algorithmic}
\usepackage{algorithm}
\usepackage{array}
\usepackage[caption=false,font=normalsize,labelfont=sf,textfont=sf]{subfig}
\usepackage{textcomp}
\usepackage{stfloats}
\usepackage{url}
\usepackage{cite}
\usepackage{verbatim}
\usepackage{graphicx}
\def\BibTeX{{\rm B\kern-.05em{\sc i\kern-.025em b}\kern-.08em
    T\kern-.1667em\lower.7ex\hbox{E}\kern-.125emX}}
\usepackage{balance}
\usepackage[colorlinks, linkcolor=red, citecolor=blue]{hyperref}

\begin{document}


\title{NOMA for Visible Light Communications:\\ Recent Advances and Future Directions}

\author{
Xuesong~Wang
\thanks{X. Wang is with the Chinese University of Hong Kong, Shenzhen, China. (e-mail: wangxuesong17@mails.ucas.ac.cn).}
}


\maketitle


\begin{abstract}

Rapidly increasing demand for high speed data is pushing 6G wireless networks to support larger link scales, lower latency, and higher spectral efficiency. Visible light communications (VLC) is a strong complement to radio frequency (RF) systems within 6G. The latest ITU G.9991 and IEEE 802.11bb standards are adapted from cable and RF wireless technologies for use in VLC, so they do not fully exploit the optical nature of light links. VLC links are often asymmetric between uplink and downlink, which makes TDMA style protocols inefficient when many users generate bursty and asymmetric traffic. Compared with RF, the strong directionality and frequent line of sight in VLC can mitigate hidden and exposed terminals, yet these effects can still appear under limited field of view, blockage, or reflections. CSMA/CA and related methods remain usable in VLC and in RF plus VLC networks, but they usually need design tweaks such as RTS/CTS or directional sensing to perform well. Although the optical spectrum is vast, the bandwidth of practical LEDs and of common PIN or APD receivers is limited, so efficient multiple access can yield large gains. This motivates a clean slate design for VLC, especially at the MAC layer. NOMA, first explored in 5G RF systems, is also promising for 6G VLC. It lets multiple users share the same time and frequency resources while tolerating controlled interference. This paper reviews progress in VLC and in NOMA based VLC, outlines key optimization constraints and objectives, surveys scenarios that fit NOMA in VLC, and points to several directions for future work.

\end{abstract}



\section{Introduction}

\begin{figure*}[htp]
\centering
\includegraphics[width=\textwidth]{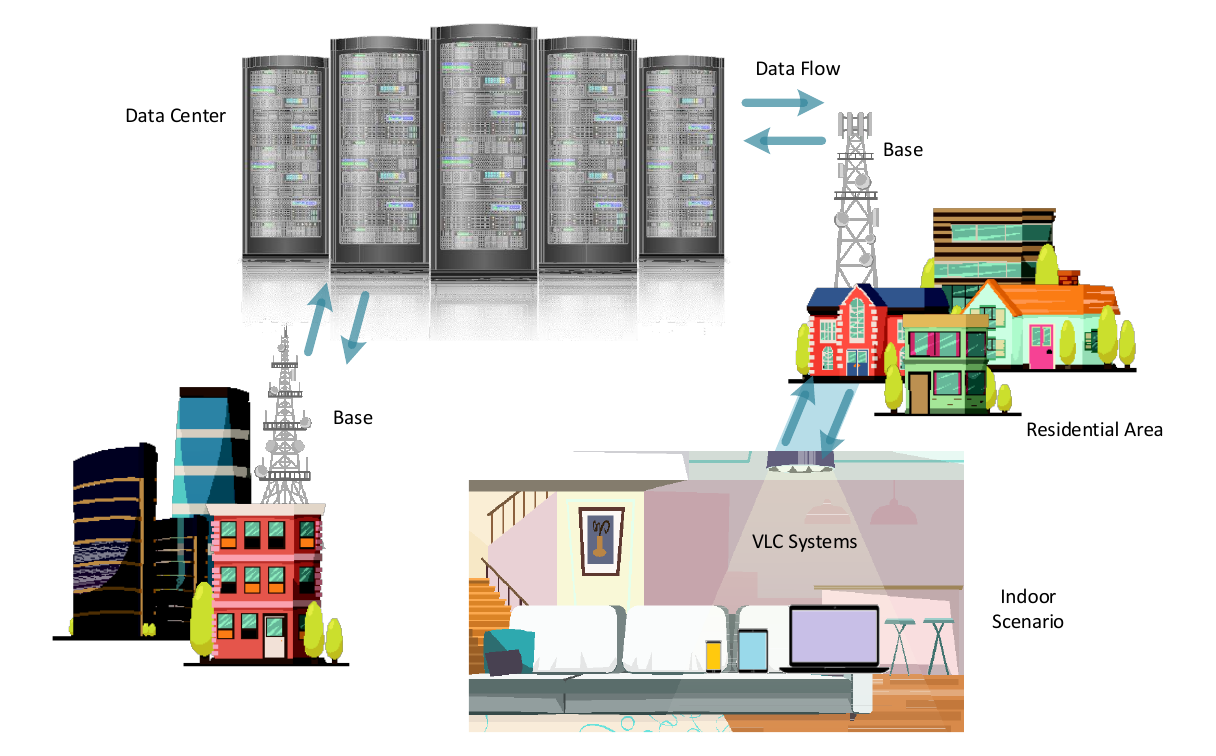}
\caption{VLC systems in 6G communication networks.}
\label{whole}
\end{figure*}

\IEEEPARstart{V}{isible} light communication (VLC) is a promising communication method in 6G that uses light-emitting diodes (LEDs) as light sources to exchange high-speed data while meeting the lighting needs (Fig. \ref{whole}). Compared with traditional lighting, LED itself has the characteristics of high efficiency, low price and long life, and its proportion in daily lighting demand is increasing year by year. Due to the ubiquity of LED light sources and the ubiquitous demand for communication, it is a natural idea to use LEDs for communication. Hence, the application of visible light communication has also developed rapidly in recent years. Considered as a technical complement to alleviate the growth pressure of wireless communication data rates by exploiting unlicensed bandwidth and cost-effective advantages \cite{Alain}, VLC systems have been studied for decades and a lot of valuable work has been done to improve the performance of VLC systems under different scenarios, such as indoor dimming optimization \cite{Zou2020}, indoor positioning \cite{Lin2017}, joint optimization of communication and lighting in unmanned aerial systems \cite{Yining2020}, non-illumination equalization with reflection \cite{Yonghe2020}, and space division multiplexing together with wavelength division multiplexing (WDM) VLC system \cite{Xinliu2021}. VLC can highly meet the system requirements of 5G or 6G with high capacity, high data rate, high spectral efficiency, high energy efficiency, low battery consumption and low latency \cite{Feng2016}.

Important challenges for visible light communications include the limited bandwidth of light-emitting diodes (LEDs), PIN diodes and avalanche photo diodes (APDs), the existence of various linear or nonlinear impairments, susceptibility to blockages, and line-of-sight (LoS) alignment requirements. In the first decade of this century, the researches on VLC could only reach the communication rate around 100 Mbps \cite{Haas2006, Le2008}. Later on, with the application of advanced pre-equalizer and post-equalizer that can stable the system performance for VLC system, as well as high spectral efficiency modulations such as carrierless amplitude and phase (CAP), pulse amplitude modulation (PAM) and discrete multi-tone (DMT) modulation, the communication rate of VLC has risen to the order of Gbps \cite{kottke, Tsonev2015}. To specify, the pre-equalizers are often used to compensate the amplitude-frequency characteristics of the LEDs at transmitters, and the post-equalizers are for impairment at receivers that are used to against the damage brought by imperfect channel.

While communication rate of researched VLC systems is much higher that can reach the vision to support the peak data rate transmission of radio frequency (RF) communications in the physical layer \cite{You2021}, there is another problem that should be put on the agenda, that is, to deal with multiple users linking to the access point (AP) efficiently under the condition when there are many users in the same domain of one AP. Usually an AP is hoped to serve multiple users at the same time. In RF communications, carrier sense multiple access with collision avoid (CSMA/CA) or time division multiple access (TDMA) are often chosen as the suitable access schemes. While the latest ITU G.9991 and the upcoming IEEE 802.11bb standards are adapted from cable and RF wireless techniques for VLC, respectively, these two access schemes could not be performed efficiently when considering VLC systems. CSMA/CA is not favourable because users typically do not sense the carrier of the other nodes in VLC systems. And, since VLC links are always asymmetric, TDMA is not efficient when there are large number of users with busy traffics. Orthogonal frequency division multiple access (OFDMA), based on orthogonal frequency division multiplexing (OFDM), has been widely researched in VLC systems. As an expansion and combination of OFDM and FDMA technique, OFDMA divides the spectrum into orthogonal pieces, then loads the transmit datas from different users on different sub-channels. Users can select sub-channels with better channel conditions for data transmission, and a group of users can access to their own certain channels at the same time. But OFDMA trade spectral efficiency for users' accessibility since the spectrum pieces cannot be fully utilized to keep them orthogonal, and LEDs, low-cost PIN diodes and APDs used in VLC system always have limited bandwidth. Non-orthogonal multiple access (NOMA), that has already been proposed in 5G, can be adapted as a more suitable choice since NOMA uses different transmission powers to transmit signals in the same frequency domain, or even the same time/code domain for further accessibility, and can provide services to multiple different users with fully utilization of channel resources at the same time. In NOMA-VLC, user can still perform OFDM as modulation type, but the channel where it is located is shared with other users. The interference introduced by this is an important basis for the receiver to use successive interference cancellation (SIC) to demultiplex signals correctly.

As a result, NOMA-based VLC systems trade increased receiver complexity for access efficiency, and the optimization of VLC system has become an important issue to tackle. Based on this fact, many papers focus on considering optimizing specific objectives under several given constraints, or enhancing NOMA-VLC systems under different scenarios. In this paper, we will first introduce the development NOMA-VLC systems, and then summarize some optimization constraints and objectives involved in the current NOMA-VLC researches. Generally, constraints and objectives are interchangeable, with one or several of them being selected as optimization objectives and the rest as constraints. Many optimization objectives or constraints have some different descriptions or alternatives, but their essences are the same or similar; moreover, some objectives or constraints overlap with each other, and we will summarize and describe them in detail. Next, we will summarize some specific scenarios targeted by the current NOMA-VLC researches with specific examples published in papers. Last but not least, we will summarize some of the cutting-edge works that can inspire our future development directions of NOMA-VLC.


\section{From OMA to NOMA}

\begin{figure*}[htp]
\centering
\includegraphics[width=\textwidth]{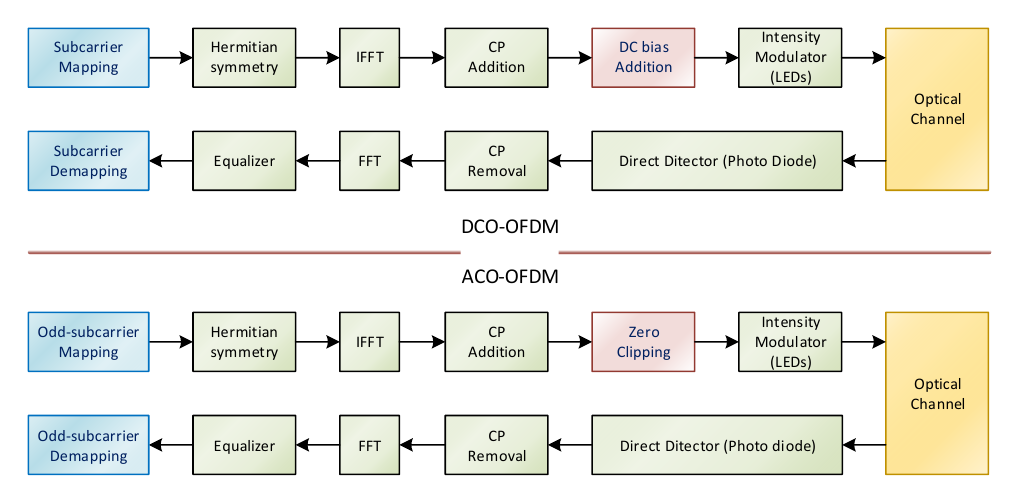}
\caption{DCO-OFDM and ACO-OFDM comparison.}
\label{DcoAco}
\end{figure*}

Multiple access (MA), that has been widely used in RF communications, is also applied in VLC to realize multiplexing of multiple users that can share resources in time domain, frequency domain or power domain. Orthogonal Multiple Access (OMA) aims to allocate a separate resource segment for each user in a certain resource domain, and the resources used by each user are orthogonal and do not interfere with each other. A specific example is OFDMA, an OFDM-based MA scheme already used in RF systems in 4G and 5G. In OFDMA, different users share the same time slot, and each user is allocated to a different frequency sub-carrier.

It is worth noting that, while OFDM or OFDMA used in the physical layer or media access control (MAC) layer of RF system are homologous, the traditional OFDM technique cannot be directly applied to VLC, mostly because the transceiver of VLC basically adopts intensity-modulated direct detection (IM-DD) as the preferred architecture. This means that the signal at the transmitter must be non-negative, whereas the RF signal under traditional MA schemes is usually complex. Several solutions have been used to address this issue, including DC-biased optical OFDM (DCO-OFDM), asymmetric clipped optical OFDM (ACO-OFDM), asymmetric clipped DC-biased optical OFDM (ADO-OFDM), and polar OFDM \cite{bawazir2018}. Among them, DCO-OFDM and ACO-OFDM are widely used and easy to implement in practical optical communication systems (Fig. \ref{DcoAco}). In DCO-OFDM, the OFDM symbol is conjugate symmetry by applying Hermitian transformation, then adds a direct current (DC) bias to ensure the signals to be positive numbers. ACO-OFDM only uses odd-subcarriers to transmit signals compared to DCO-OFDM, with no DC bias added but directly clipped the signal from zero. Some of the main differences that exist between the two methods are summarized below.

\begin{itemize}
    \item ACO-OFDM needs a negative clipping operation before the D/A conversion at the transmitter, so that the values on the odd-numbered subcarriers become half of the original values in the frequency domain.
    \item The spectrum utilization of ACO-OFDM is half of that in DCO-OFDM.
    \item In ACO-OFDM, the time-domain signals are all positive real numbers, so only a small DC offset is required to ensure the LED's luminescence, while DCO-OFDM needs a larger DC offset that reduces power efficiency.
\end{itemize}

As mentioned above, when the uplink demand becomes intensive, the limited spectrum resources of LEDs, PIN diodes and APDs may not be able to meet the simultaneous access requirements of a large number of users. Therefore, NOMA has been widely studied as a new MA method in VLC system. NOMA is a power-domain multiplexing technique that allows users to transmit signals at different power levels simultaneously. The OFDM can still be used in NOMA to achieve higher receiver sensitivity \cite{You2021}, and the sub-channels in a certain resource domain (for example, the frequency domain) can still be kept orthogonal, but one sub-channel is allocated to multiple users. The receiver needs to perform SIC to separate the signals from each other. SIC eliminates interference iteratively by demultiplexing the transmitted signal one by one in the received superimposed signals according to their different receiving powers, then striping it off and repeating the steps. However, the channel state always changes during the transmission, the receiver has to detect the received powers and sort them in order dynamically. As a result, the receiver in NOMA-VLC system is much more complex than that in OMA-VLC system.

There are some other characteristics of NOMA-VLC system.
\begin{itemize}
    \item Although the channel is always noisy, for example, thermal and shot noise are often considered in VLC communications, the received signal-to-noise ratio (SNR) of VLC is generally higher than RF, and using NOMA with SIC can benefit from this.
    \item The system has better adaptability to link changes because the access point in NOMA-VLC system may not rely on the channel state information (CSI) fed back by users.
    \item System optimization for NOMA VLC can always be described as a non-convex optimization problem. Since non-convex problems are difficult to solve, an approximate convex description of the original non-convex formulations is often an important step in many papers.
\end{itemize}

\begin{figure}[tp]
\centering
\includegraphics[width=\linewidth]{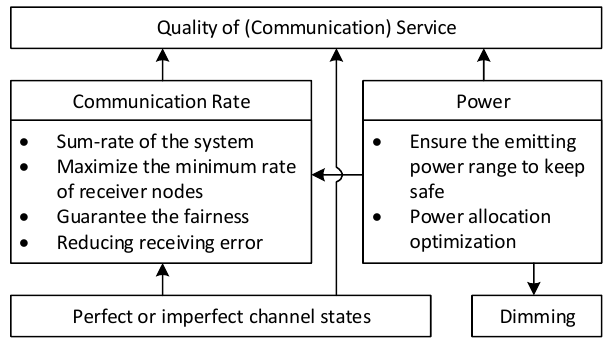}
\caption{Constraints and their relationships.}
\label{Contraints}
\end{figure}


\section{Constraints and Targets}

Researches on NOMA-VLC systems is often abstracted as an optimization problem in a certain scenario. Generally, the essence of an optimization problem is to select a set of variables or parameters so that the objective function (including decision variables) reaches the optimal value under a series of related constraints (equality constraints or inequality constraints). The optimal value can be either maximization in a positive sense or minimization in a negative sense. Some of the factors frequently considered in NOMA-VLC systems are summarized below. Often these factors can be constraints or goals of each other, i.e. one or a few of them are constraints and the rest are goals. Each factor may also contain some variation, or there may be some parameters overlap between some factors (Fig. \ref{Contraints}).

\subsection{Communication rate}

The communication rate has always been the most concerned issue of the communication system. Regardless of the factors involved in the optimization, it needs to ensure that the communication rate itself is acceptable. There are many variants of this factor, such as the sum-rate of all users at the receiver side \cite{FeifanYang2021}, maximizing the minimum user rate to avoid the barrel effect \cite{Shen2017}, or the fairness of the rate per receiver \cite{Raj2022, Yang2016}. From another perspective, the communication rate and the error rate are negatively correlated, and many studies also focus on reducing errors \cite{marshoud2017, Jindal2020}. In \cite{Tahira2019}, the authors abstracted the two key parameters, system capacity and received power, that directly affect the received signal-to-noise ratio, as a non-convex problem. They mainly focused on the downlink of the NOMA-VLC system, and considered several constraints such as total power and single-user power allocation, as well as channel parameters such as LED’s semi-angle at half-power, field of view (FOV), and filter gain. When the received SNR is 8dB, the optimized BER results is one order less than traditional scheme of both near or far users.

Communication rate can directly affect the quality of service when higher layers of the communication systems, such as network layer or application layer, need a high-speed and stable low-layer link, and may be influenced by channel states or power allocation of the NOMA receiver. And it is often considered in the downlink of the NOMA-VLC system since the users' performances are easy to follow and can reflect the system's performance.

\subsection{Power and energy efficiency}

The power involved in NOMA-VLC can be divided into two parts: transmit power and receive power. Considering the power safety of the transmitter and the user's eye safety under the LED, or in order to expand the LED working area, the emission power should be located within a suitable range. Optimizing the power of the transmitter can also improve energy efficiency. In NOMA, different power levels are assigned to different users, where data is multiplexed in the power domain using superposition coding at the transmitter, enabling each user to utilize the full available bandwidth at any time. Given that SIC is one of the most common ways of demultiplexing a signal at the receiver, it is critical to allocate the appropriate transmit power to users. Users with poor channel conditions will allocate more power to solve the flaws caused by the near-far effect and make the system performance uniform.

There are several methods of allocating power among different transmitters, such as:
\begin{itemize}
    \item static power allocation (SPA) \cite{Marshoud2018}, where the power allocation factor is independent of the user's channel conditions;
    \item gain ratio power allocation (GRPA) \cite{Marshoud2015}, by allocating power for users according to optical channel gain;
    \item normalized gain difference power allocation (NGDPA) \cite{Chen2017} that is better than GRPA in $2\times2$ multiple-input multiple-output (MIMO) scheme;
    \item fair power allocation \cite{Jindal2020}, where power is allocated such that the observed BERs remain consistent at each user.
\end{itemize}

In \cite{Wang2021}, considering the practical requirements of indoor VLC systems such as severe high frequency fading due to limited bandwidth and signal peak power limitation due to eye safety and non-restriction, the authors proposed an efficient joint subcarrier and power allocation algorithm that leverages a logarithmic utility function of the capacity of each user in each subcarrier to balance throughput and user fairness in each subcarrier.

\subsection{Quality of service}

Quality of service (QoS) considered in most researches includes the coordination of the QoS of each users within the system, such as ensuring that users in the system far from the transmitter LED can effectively receive information; and the overall quality of service of the system, such as the whole system's throughput \cite{Yin2016}. Because of the near-far effect of NOMA-VLC mentioned above, although the performance of edge users is relatively limited, the total user throughput of the system and the throughput of edge users are equally important to ensure the quality of service. Sometimes the minimum communication rate of each node needs to be above a certain threshold to guarantee the QoS. The derived results are often simulated and demonstrated using Monte Carlo simulations.

In addition, since the received power is also related to the communication rate, power allocation can also affect the quality of service directly. To ensure QoS of NOMA-VLC system, it can be achieved by optimizing transmitter power allocation or reducing the power allocation complexity.

\subsection{Perfect or imperfect channel states}

When implementing a NOMA-VLC system, channel state information (CSI) is prioritized as a known factor in some papers. But CSI may not always be perfect, and noisy CSI can degrade system performance. In addition, rapid changes in the channel caused by user movement may result in untimely CSI updates, which leads to adverse performance losses. Monte Carlo simulations can be used to study the error rate performance of NOMA-VLC system with perfect or imperfect CSI that may be noisy or stale \cite{marshoud2017}, and approximate solutions of the simulated results can be representative and persuasive. Also the closed-form expression of average BER can be derived by assuming some ideal conditions \cite{Tran2022}, but the result may not be universal.

\subsection{Dimming}

Unlike traditional RF communications, stable lighting itself is an important factor in VLC systems since it is the essence of using LEDs. In order to ensure the dimming capability of the transmitter, the ratio of 0/1 in the transmit codeword should be considered roughly evenly proportioned, or the receiver can use multiple LEDs for beamforming \cite{Zou2020}. In \cite{Cang2022}, a power allocation study considering Gaussian noise (thermal) and shot noise was conducted, where the optimization objective was to maximize the minimum value of signal to interference plus noise ratio (SINR) since the closed-form of shot noise was intractable, and guaranteed a wide range of dimming that is safety for eyes and fairness for users. The study in \cite{marshoud2017} considered the BER performances of two dimming schemes, called OOK analog dimming and variable OOK (VOOK) dimming, that could be supported in NOMA-VLC systems.

\begin{figure*}[htp]
\centering
\includegraphics[width=\textwidth]{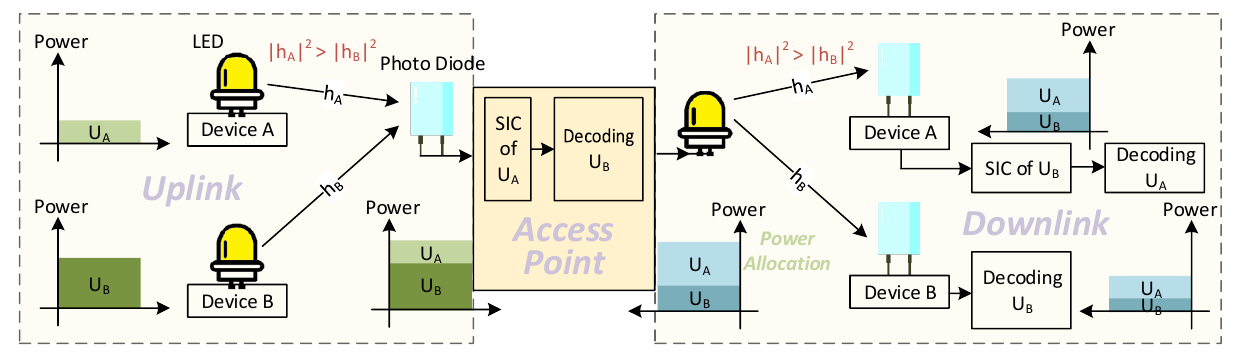}
\caption{NOMA-VLC system with uplink and downlink. Note that both "device A" in uplink and in downlink are the same device; so is the device B.}
\label{NomaVlc}
\end{figure*}


\section{Different Scenarios and Features in NOMA-based VLC systems}

Optimization problems must be accompanied by specific scenarios. Since lighting systems are ubiquitous, the scene requirements for visible light communication can also be considered ubiquitous. Considering the characteristics of visible light communication and some specific scenarios that already exist in the radio frequency system, some typical scenarios involved in the current NOMA-VLC systems are summarized as follows.

\subsection{Multiple-input multiple-output systems}

Multiple-antennas is one of the main structures to improve spectral efficiency. Compared to massive MIMO systems that are the core technology of 5G, ultra-massive MIMO systems with larger arrays are expected to double the spectral efficiency of 6G. For VLC systems, massive MIMO means using LED arrays to transmit signals, and those are ubiquitous in today's lighting systems. The paper \cite{Dixit2022} analyzed the bit error rate performance of all users on a LoS channel in a multi-pulse position modulation (M-PPM) modulated MIMO-based multi-user NOMA-VLC system. Considering the actual situation of imperfect SIC, the authors deduced a closed-form for the error probability of the system. They also included a comprehensive analysis of two-user and three-user M-PPM modulated $2\times2$ MIMO-NOMA-VLC system and derived accurate error probability expressions. As the number of photodetectors per user increases, the error probability of the considered system is reduced.

\subsection{Underwater VLC systems}

Underwater VLC typically operates in the blue-green window (approximately 450–550~nm) to exploit the minimum absorption band of seawater. Since seawater exhibits good conductor properties for high-frequency electromagnetic waves, the skin effect generated limits the communication distance of radio frequency communication underwater, while the link of underwater visible light communication can still be established. In this case, VLC is treated as a viable solution to enable gigabit wireless network underwater to extend the capability beyond terrestrial cellular communications \cite{You2021}. Underwater optical communication mainly uses the blue-green light band (400 to 600 nm), because the absorption of this band by seawater is much lower than that of other bands. Turbulence, bubbles, scintillation, turbidity are important considerations in the underwater channel. In \cite{Bariah2021}, a new outage probability analytical expression was derived to quantify the performance of NOMA under different turbulence scenarios, taking into account the path loss modeled as a semi-collimated laser source with a Gaussian beam shape, and the turbulence modeled using log-normal distribution with additive white Gaussian noise (AWGN).

\subsection{Security issues in VLC systems}

The security of the VLC system mainly considers two aspects: power limit eye safety for normal users, and protection against eavesdropping by other unfriendly users. As mentioned above, authors in \cite{Cang2022} considered the eye safety by adopting a wide-range dimming ability. In \cite{Peng2021}, considering the case of maximum ratio combining (MRC) based on signal detection adopted by eavesdroppers, the paper designed pre-coding to provide sufficient protection for legitimate transmissions. For the case where the location of the eavesdropper is known, they derived the secrecy capacity by considering the IM-DD of the VLC system, and formulated the secrecy capacity optimization under several issues such as the constraints of SIC order, perfect SIC condition, total transmit power and maximum allowable signal amplitude.

The security of underwater VLC communication was also considered in \cite{Kumar2022}. This paper proposed two optimal LED selection (OLS) and sub-optimal LED selection (SLS) schemes based on NOMA to determine the selection of emitting LEDs, so as to ensure the highest security rate and transmit information, and prevent passive eavesdropping attacks. Meanwhile, the received signal fluctuations were modeled using an exponential generalized gamma distribution to measure underwater turbulence effects, as it can provide excellent agreement between theoretical and experimental results.

\subsection{Backhaul link hybridising with VLC}

Radio over fiber (RoF) can provide large capacity as the backbone transmission link of high-speed, cost-effective communication network \cite{Rahman2020}. In \cite{Manh2021}, an optical backhaul cooperative NOMA with VLC system was proposed. The authors considered hybrid cooperative three-hop optical FSO/RoF/VLC system where FSO/RoF nodes are treated as backhaul link to an indoor VLC system that contains two local users. The source node needs to communicate with the users in VLC system through an FSO link that is positioned on top of two high-rises, and an RoF link is employed as a relay to amplify and forward signals from one head of FSO link to the VLC system. They considered the nonlinear effects of the backhaul link, and simulated the VLC system using Gamma-Gamma non-fading channel. Average BERs of BPSK modulation of users and systems were evaluated after the derivation of cumulative distribution function (CDF) expression of equivalent end-to-end SNR of two users. Simulation results showed that high-speed optical backhaul link with NOMA can achieve great performance improvement.

\subsection{Expansion to higher modulation formats}

Higher order modulations such as 16-quadrature amplitude modulation (QAM) and 64-QAM can be used to improve spectral efficiency. However, due to the complexity of the constellation set, the traditional SIC method is difficult to decode the superimposed signals. In \cite{Ren2018}, authors described a method called ergodicity and comparison (EAC) to decode superimposed signals under higher order modulations. EAC can recover signals received from both users simultaneously rather than decoding them one by one that SIC normally does, by specifying a chosen trail constellation values from one user and iteratively calculating the difference between noisy received values with the ones in original constellation set from the other user to find the minimum distance. With this method, a two-user NOMA scheme with up to 64-QAM is simulated to get better results than normal SIC.


\section{Future Works in NOMA-VLC System}

In this section, we will summarize some of the more cutting-edge work that we think could inspire the next development direction for NOMA-VLC.

\begin{figure*}[htp]
\centering
\includegraphics[width=\textwidth]{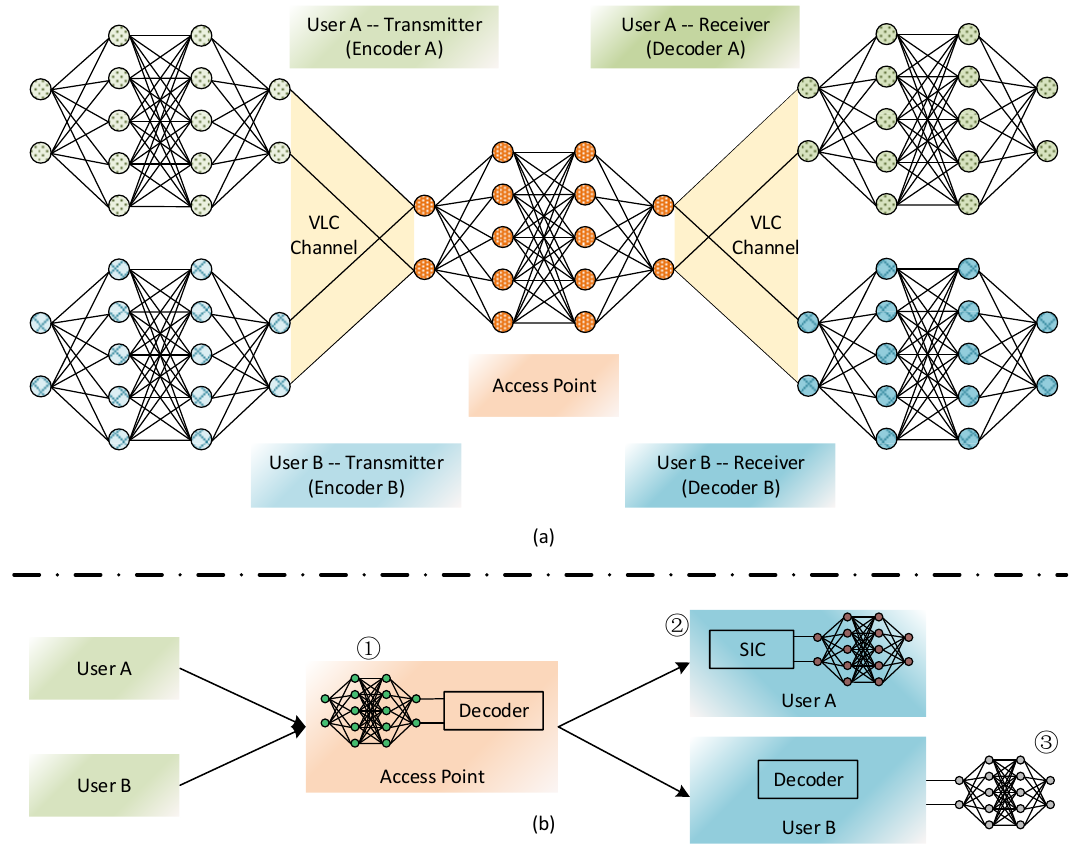}
\caption{Machine learning based NOMA-VLC system scheme. There are two major aspects: (a) DNN-based full-link NOMA VLC system model, and (b) use neural network to replace some modules of the system, such as SIC module, decoding module, or data processing module.}
\label{dnn}
\end{figure*}

\subsection{New dimension of multiple access collaborated NOMA scheme}

NOMA only arranges transmit signals' superimposing and performs corresponded receiver algorithms in the power domain, and does not have limitations in other domains. It inspires us that NOMA can be combined with multiple access schemes in other domains to further enhance the utilization of limited resources.

Sparse Code Multiple Access (SCMA) encodes transmitted data using multidimensional codewords selected from a set of predefined codebooks, and the receiver performs a message passing algorithm (MPA) to detect the multiplexed codewords. This decoding method is inspired by the field of codecs, where MPA is commonly used in graph-based codec systems. In \cite{Bang2019}, the authors proposed a NOMA scheme called power domain sparse code multiple access (PD-SCMA) for VLC systems, which uses both the power domain and the code domain for multi-user signal transmission on subcarriers. The feasibility of PD-SCMA VLC is verified by experiments with two power-domain multiplexing layers, which can provide twice the total data rate than SCMA.

In \cite{Song2021}, authors proposed a novel MA scheme that users in VLC systems adopt different kind of modulations. The authors elaborated a VLC system containing two users that the one with higher-data-rate request performs M-PPM, while the other with lower-data-rate request simply performs on-off keying (OOK). Intuitively, the receiver has significantly different power levels for the two signals, and it is easy to distinguish both signals from the superimposed one like multiple user detection (MUD) does in wireless communication systems.

In \cite{mitra2020}, the authors introduced the use of multi-color LEDs as emitters, in which several users use SCMA to overlap on each color. Therefore, the SCMA codewords on each color overlap in the color domain before transmission. At the receiver, the residual color coupling is removed by inverting the coupling matrix after color filtering. Finally, the MPA is used to recover the user signal on each color channel. In addition, asymptotic error rate expressions were derived for the proposed color-domain SCMA.

\subsection{SIC-free NOMA-VLC systems}

As the common algorithm at the receivers in NOMA-VLC systems, SIC increases the complexity of the receivers. First, the received signals are sorted according to the signal power. In the SIC signal detection process, a very important point is the order of user detection, and the sorting is performed here according to the signal power of the user. In NOMA, the transmitter will use power multiplexing to allocate power to different users, and the order of users according to the received signal power determines the best receiving effect (Fig. \ref{NomaVlc}). In the actual process, the power of users is constantly changing. This requires the SIC receiver to continuously sequence user power to avoid error propagation.

Due to the complexity of SIC, there are several researches focused on scenarios that are carefully calibrated that do not need SIC at the receivers. In \cite{Zhang2022}, a multi-user superposition transmission based SIC-Free NOMA-VLC system was proposed. Constellations from both users at the transmitter are mapped into a composite constellation by linear and nonlinear superposition coding. In \cite{Chen2018}, a new NOMA scheme based on constellation partitioning coding (CPC) and uneven constellation demapping (UCD) is implemented for downlink VLC systems. By using CPC/UCD, user decoding can be achieved without SIC, thus eliminating the effect of error propagation caused by imperfect SIC, thereby improving BER performance. In addition, by choosing an appropriate bit allocation scheme, flexible rate multiple access can also be achieved.

\subsection{Combination with machine learning}

Machine learning (ML) is quite hot in dealing with issues in communication systems recently, since ML is hoped to achieve continuous information sharing and fully immersive persistent connectivity for humans, machines and massive devices in 6G. A typical neural network based NOMA-VLC system can be constructed as shown in Fig. \ref{dnn} that learned from wireless communications. Many recent researches treat ML as equalizer or demodulator in RF or VLC systems, but only a few papers involved ML in NOMA-based RF or VLC systems. In \cite{Lin2021}, a NOMA-VLC signal demodulator based on convolutional neural network (CNN) was proposed, where signal compensation and recovery are jointly implemented. This CNN architecture can mitigate linear and nonlinear distortions in NOMA-VLC to improve system performance. The captured NOMA signal is used to train the CNN in offline mode, and directly applied to a CNN-based demodulator for signal compensation and recovery in online mode. Also in \cite{Manh2021}, a deep learning based framework was proposed to accurately estimate the traversal capability of the proposed system and can reduce the complexity cost of theoretical analysis markedly. We believe that machine learning still has huge potential to be fully utilized in wireless communication systems, including NOMA-VLC.

\section{Conclusion}

As one of the promising communication candidate for next-generation wireless communication networks, visible light communication (VLC) is an important complement to the radio frequency (RF) communications. However, limited spectrum of VLC devices and imperfect VLC links mitigate the system performance, while multiple access (MA) schemes such as TDMA or CSMA/CA are not efficient or suitable. As the first-proposed technique in 5G, non-orthogonal multiple access (NOMA) can be used by allowing multiple users to utilize the available time and frequency resources simultaneously, and is also suitable for the VLC communications in 6G. Since NOMA improves the accessibility of the visible light communication system with signals' interference at the receiver, the optimization of NOMA-VLC system becomes an considerable problem to deal with the decrease of reliability and the increase of system complexity. In this paper, we introduce some issues when considering NOMA in VLC systems, and present the optimization goals and constraints involved in current related researches, then summarize some detailed application scenarios. We believe that NOMA will become an effective technology for realizing VLC system with more researches in the future, and further exert its utility with some state-of-the-art techniques.

\bibliographystyle{IEEEtran}
\bibliography{references.bib}


\end{document}